\documentclass[11pt,oneside,english]{amsart}
\usepackage{mathpazo}
\usepackage[scaled=0.92]{helvet}
\usepackage{courier}
\usepackage[T1]{fontenc}
\usepackage[latin9]{inputenc}
\usepackage[a4paper]{geometry}
\usepackage{verbatim}
\usepackage{amsbsy}
\usepackage{amstext}
\usepackage{amsthm}
\usepackage{stackrel}
\usepackage{graphicx}
\usepackage{setspace}
\usepackage[authoryear]{natbib}
\makeatletter

\numberwithin{equation}{section}
\numberwithin{figure}{section}

\theoremstyle{plain}

\numberwithin{equation}{section}
\usepackage{tikz}
\makeatother

\usepackage{babel}
\begin{document}

\title[Non-Central Chi Regression]{Bayesian Non-Central Chi Regression for Neuroimaging}

\author{Bertil Wegmann, Anders Eklund and Mattias Villani}

\thanks{Wegmann: \textit{Division of Statistics and Machine Learning, Dept.
of Computer and Information Science, Linköping University, SE-581
83 Linkoping, Sweden}. \textit{E-mail: bertil.wegmann@liu.se}. Eklund:
\textit{Division of Statistics and Machine Learning, Dept. of Computer
and Information Science, Linköping University, SE-581 83 Linköping,
Sweden}. \textit{Division of Medical Informatics, Dept. of Biomedical
Engineering, Linköping University, SE-581 85 Linköping, Sweden. Center
for Medical Image Science and Visualization (CMIV), Linköping University,
Linköping, Sweden. E-mail: anders.eklund@liu.se}. Villani: \textit{Division
of Statistics and Machine Learning, Dept. of Computer and Information
Science, Linköping University, SE-581 83 Linköping, Sweden}. \textit{E-mail:
mattias.villani@liu.se}.}
\begin{abstract}
We propose a regression model for non-central $\chi$ (NC-$\chi$)
distributed functional magnetic resonance imaging (fMRI) and diffusion
weighted imaging (DWI) data, with the heteroscedastic Rician regression
model as a prominent special case. The model allows both parameters
in the NC-$\chi$ distribution to be linked to explanatory variables,
with the relevant covariates automatically chosen by Bayesian variable
selection. A highly efficient Markov chain Monte Carlo (MCMC) algorithm
is proposed for simulating from the joint Bayesian posterior distribution
of all model parameters and the binary covariate selection indicators.
Simulated fMRI data is used to demonstrate that the Rician model is
able to localize brain activity much more accurately than the traditionally
used Gaussian model at low signal-to-noise ratios. Using a diffusion
dataset from the Human Connectome Project, it is also shown that the
commonly used approximate Gaussian noise model underestimates the
mean diffusivity (MD) and the fractional anisotropy (FA) in the single-diffusion
tensor model compared to the theoretically correct Rician model.\vspace{0.1cm}
\end{abstract}

\maketitle

\section{Introduction}

Gaussian statistical models are very common in the field of neuroimaging,
as they enable efficient algorithms for estimation of brain activity
and connectivity. However, the measured signal in diffusion weighted
imaging (DWI) and functional magnetic resonance imaging (fMRI) is
the magnitude of a complex-valued Gaussian signal and therefore follows
a Rician distribution, see \citet{gudbjartsson1995rician} and Section
\ref{subsec:Rician-regression}. The Gaussian model is a good approximation
to the Rician model in fMRI as the signal-to-noise (SNR), defined
here as the ratio of the average BOLD signal to its standard deviation,
for fMRI data tends to be large enough for the approximation to be
accurate \citep{AdrianMaitraRowe2013Ricean}. However, the recent
push towards higher temporal and spatial resolution in neuroimaging
\citep{Moeller,feinberg2012rapid,Setsompop} may lead to low SNRs
with increased risk of distorted conclusions about brain activity
and connectivity. This is demonstrated in Section \ref{sec:ActivityfMRI},
where a Rician model is able to accurately detect brain activity at
low SNRs, while the Gaussian approach fails to do so. Low SNRs are
also common for DWI , especially when the b-value is high \citep{ZhuEtAl2009JASARiceRegr}.
Using a Gaussian model for diffusion tensor imaging (DTI) can therefore
lead to severely misleading inferences. The reason for the popularity
of the Gaussian approach is that Gaussian models can be analyzed using
simple algorithms, while the Rician distribution is complicated since
it does not belong to the exponential family. More generally, MR images
collected by simultaneous acquisition from $L$ independent coils
may follow the non-central $\chi$ (NC-$\chi$) distribution with
$L$ degrees of freedom, depending on how the measurements are combined
into a single image \citep{tristan2012least,fernandez2016}. We therefore
derive our algorithm for the general NC-$\chi$ model from which the
Rician model can be directly obtained as the special case when $L=1$.

\subsection{Rician models in fMRI}

There have been a handful of approaches for the Rician model in fMRI
applications. \citet{solo2007EMRician} and \citet{ZhuEtAl2009JASARiceRegr}
propose to augment each data observation with the missing phase information,
and to use the EM algorithm to obtain the maximum likelihood estimates
of the regression coefficients in the Rician model; \citet{AdrianMaitraRowe2013Ricean}
provide the extension to the case with autocorrelated errors. Although
not discussed in the literature, the data augmention technique is
naturally extended to a fully Bayesian analysis via Gibbs sampling,
where the parameters are iteratively sampled conditional on the missing
phase observations, followed by a sampling step for the phases given
the model parameters. The convenience of introducing unobserved phase
information does not come without cost, however, and data augmentation
is well known to lead to inefficient exploration of the posterior
distribution and inflated numerical standard errors \citep{liu1994covariance}.
The same problems tend to plague the EM algorithm, which often exhibit
very slow convergence.

\subsection{Rician models in DTI}

Rician models have mainly been used for noise removal in DTI \citep{Basu2006,Wiest,Fernandez},
but also for tensor estimation \citep{andersson,Veraart}. The only
method that we are aware of for estimating diffusion parameters in
the more general NC-$\chi$ regression model, for data acquired with
several independent coils, is the (non-Bayesian) least squares approach
presented by \citet{tristan2012least}.

\subsection{Non-central chi regression}

We therefore introduce a NC-$\chi$ regression model where both parameters
in the distribution (the mean and variance of the underlying complex-valued
signal) are modeled as functions of covariates, with the Rician model
as an important special case. We propose a Bayesian analysis of the
model based on a highly efficient Markov Chain Monte Carlo (MCMC)
algorithm, to simulate from the joint posterior distribution of all
model parameters. Contrary to previous Bayesian and EM approach, our
Bayesian methods works directly on NC-$\chi$ or Rician distributions,
without the need to introduce missing phase data, and the MCMC convergence
is excellent due to an accurately tailored proposal distribution.
A high efficiency makes it possible to use a smaller number of simulations
to obtain the same numerical accuracy. This is absolutely crucial
for imaging applications since a separate MCMC chain is run for each
voxel. Moreover, our MCMC algorithm also performs Bayesian variable
selection among both sets of covariates. For both DTI and fMRI data,
our Bayesian approach has the obvious advantage of capturing the uncertainty
in each voxel. The uncertainty can easily be propagated to the group
analysis, to down-weight subjects with a higher uncertainty. This
is in contrast to the popular TBSS approach (tract-based spatial statistics)
\citep{Smith} for voxel-wise multi-subject analysis of fractional
anisotropy (FA), which ignores the uncertainty of the FA.

Using a freely available DWI dataset from the Human Connectome Project
\citep{VanEssen}, we show that commonly used Gaussian DTI approximation
underestimates the mean diffusivity (MD) and substantially underestimates
the FA of the single-diffusion tensors, compared to the theoretically
motivated Rician model, especially in white-matter regions with high
FA. In addition, we show that covariates are needed in both parameters
of the Rician distribution, not only in the mean. In an fMRI simulation
study, we formulate a sensible prior distribution for the regression
coefficients based on the Fisher information matrix, and demonstrate
that the Rician model is remarkably adept at recovering the activations
even at very low SNRs. We also show that the Gaussian model is likely
to lead to severely erroneous activation inference in such settings.

\subsection{Application to more advanced diffusion models}

We have here focused on the rather simple single-diffusion tensor,
while more recent work focus on extending the diffusion tensor to
higher orders. In the work by \citet{Westin}, a regression approach
is used to estimate the diffusion tensor and a fourth order covariance
matrix in every voxel. Our regression framework can therefore easily
be applied to QTI (q-space trajectory imaging) data \citep{Westin}
as well, and more generally for any diffusion model that can be estimated
using regression. Moreover, DTI is still the most common choice for
studies investigating FA differences between healthy controls and
subjects with some disease \citep{Shenton,Eierud}.

\section{Heteroscedastic Rician and NC-$\chi$ Regression}

We start by describing our model for the special case of a Rician
distribution, and then generalize it to the NC-$\chi$ case.

\subsection{Rician regression\label{subsec:Rician-regression}}

The measured signal in DTI and fMRI is a complex-valued indirect measure
of structural brain connectivity and brain activity, respectively,
\[
\tilde{y}_{t}=a_{t}+b_{t}\cdot i,
\]
where the real part $a_{t}\sim N\left(\mu_{t}\cos\theta_{t},\phi_{t}\right)$
and the imaginary part $b_{t}\sim N\left(\mu_{t}\sin\theta_{t},\phi_{t}\right)$
are independent, and the mean
\[
\ln\mu_{t}=\beta_{0}+\mathbf{x}_{t}'\beta
\]
is a linear function of a vector of covariates $\mathbf{x}_{t}$ at
measurement $t$. In fMRI the vector $\mathbf{x}_{t}$ typically contains
the stimulus of the experiment convolved with a hemodynamic response
function, polynomial time trends and head motion parameters, while
$\mathbf{x}_{t}$ mainly contains gradient directions in DTI. Note
that $\phi_{t}$ is potentially measurement-varying, to allow for
heteroscedastic complex-valued noise.

It is rare to analyze the complex signal measurements $a_{t}$ and
$b_{t}$ directly (\citep{RoweLogan2004} and follow-up papers are
exceptions). The most common approach is to use the magnitude of $\tilde{y}_{t}$
as response variable, i.e.
\[
y_{t}=\left|\tilde{y}_{t}\right|=\sqrt{a_{t}^{2}+b_{t}^{2}}.
\]
It is well-known that the magnitude follows a Rician distribution
\citep{rice1945mathematical} with density function
\[
p(y|\mu,\phi)=\frac{y}{\phi}\exp\left(-\frac{\left(y^{2}+\mu^{2}\right)}{2\phi}\right)I_{0}\left(\frac{y\mu}{\phi}\right),
\]
for $y>0$ and zero otherwise. 

The discussion above uses $t$, as in time, as subscripts for the
observations. This is suitable for fMRI time series, but to emphasize
that our models can also be used for DWI data (see Section \ref{sec:DWI}),
we will in the remainder of the paper use the more generic $i$ to
denote observations. We propose the following heteroscedastic Rician
regression model
\begin{align}
y_{i}|x_{i},z_{i},\mu_{i},\phi_{i} & \sim Rice(\mu_{i},\phi_{i})\;\text{ for }i=1,...,n,\nonumber \\
\ln\mu_{i} & =\beta_{0}+\mathbf{x}_{i}^{T}\beta,\nonumber \\
\ln\phi_{i} & =\alpha_{0}+\mathbf{z}_{i}^{T}\alpha,\label{eq:RiceModel}
\end{align}
and independence of the $y_{i}$ \textit{conditional} on the covariates
in $\mathbf{x}_{i}$ and $\mathbf{z}_{i}$. Since $\mathbf{x}_{i}$
and $\mathbf{z}_{i}$ may contain lags of the response variables,
our model can capture temporal dependence in fMRI. Note also that
we allow for heteroscedasticity in the complex signal, since the variance
of the underlying complex-valued signal $\phi_{i}$ is a function
of the regressors in $\mathbf{z}_{i}$. Although the model in Eq.
\ref{eq:RiceModel} has the same structure as a generalized linear
model (GLM) \citep{mccullagh1989generalized}, it is actually outside
the GLM class since the Rician distribution does not belong to the
exponential family. The logarithmic link functions used in Eq. \ref{eq:RiceModel}
can be replaced by any twice-differentiable invertible link function.

\subsection{NC-$\chi$ regression}

Both fMRI and DWI images may be obtained from parallel acquisition
protocols with multiple coils, often used to increase the temporal
and spatial resolution. Under the assumption of independent complex
Gaussian distributed noise in each coil, the sum of squared magnitudes
follow the non-central $\chi$ (NC-$\chi$) distribution \citep{tristan2012least,fernandez2016}.
The non-central $\chi$ density with $2L$ degrees of freedom is of
the form 
\begin{equation}
p(y|\mu,\phi,L)=\mbox{\ensuremath{\frac{y^{L}}{\phi\mu^{L-1}}}}\exp\left(-\frac{y^{2}+\mu^{2}}{2\phi}\right)I_{L-1}\left(\frac{y\mu}{\phi}\right),\label{eq:NCdensity}
\end{equation}
for $y,\mu,\phi>0$. We denote this as $y\sim\,$NC-$\chi$. Note
that when $L=1$, the density in Eq. \ref{eq:NCdensity} reduces to
the $\mathrm{Rice}(\mu,\phi)$ density. Similarly to the Rician case,
we can model $\mu$ and $\phi$ as functions of explanatory variables
via logarithmic link functions. In summary, the observations are assumed
to be independently NC-$\chi$ distributed conditional on the explanatory
variables, according to 
\begin{align}
y_{i}|x_{i},z_{i} & \sim NC-\chi(\mu_{i},\phi_{i},L)\nonumber \\
\ln\mu_{i} & =\beta_{0}+\mathbf{x}_{i}^{T}\beta,\nonumber \\
\ln\phi_{i} & =\alpha_{0}+\mathbf{z}_{i}^{T}\alpha.\label{eq:NC-model}
\end{align}
Lagged response values may again be used as covariates in $\mu$ and
$\phi$ to induce temporal dependence.

The order $L$ of the NC-$\chi$ distribution may be given by the
problem at hand, for example by the number of independent coils used
for data collection. Due to the lack of perfect independence between
coils and other imperfections, $L$ is often unknown and needs to
be estimated from the data. Note that $L$ can in general be any positive
real number in the NC-$\chi$ distribution, and does not need to be
an integer. Our approach makes it straightforward to introduce an
MCMC updating step, to simulate from the conditional posterior distribution
of $\ln L$, or even model $\ln L$ as a linear function of covariates.
\begin{comment}
Jag, BW, har hittat hur man beräknar $\frac{\partial I_{L-1}(z)}{\partial L}$,
men endast för heltal $L$. Dessutom är derivatan rätt så stökig och
det blir ännu värre för andraderivatan. Jag tycker därför att vi kan
skippa att ta med detta i Appendix, eftersom vi också inte använder
oss av NC-chi och det därför räcker med att beskriva i texten ovan
att vi kan lägga till ett Gibbs-steg i MCMC:n.

MV: Ok.
\end{comment}

\section{Bayesian Inference}

The Bayesian approach formulates a prior distribution for all model
parameters, and then updates this prior distribution with observed
data through the likelihood function to a posterior distribution. 

\subsection{Posterior distribution and posterior probability maps}

The aim of a Bayesian analysis is the joint posterior distribution
of all model parameters
\[
p(\beta,\alpha\vert\mathbf{y},\mathbf{X},\mathbf{Z})\propto p(\mathbf{y}\vert\beta,\alpha,\mathbf{X},\mathbf{Z})p(\beta,\alpha),
\]
where $p(\mathbf{y}\vert\beta,\alpha,\mathbf{X},\mathbf{Z})$ is the
likelihood function for the MR signal, $p(\beta,\alpha)$ is the prior,
$\mathbf{y}=(y_{i})_{i=1}^{n}$, $\mathbf{X}=(\mathbf{x}_{i}^{T})_{i=1}^{n}$
and $\mathbf{Z}=(\mathbf{z}_{i}^{T})_{i=1}^{n}$; we are here including
the intercepts in $\beta$ and $\alpha$. Based on this joint posterior
one can compute the marginal posterior of any quantity of interest.
From the joint posterior $p(\beta,\alpha\vert\mathbf{y},\mathbf{x},\mathbf{z})$,
it is straight forward to compute Posterior Probability Maps (PPMs),
see \citet{FristonPenny2003PPM}. For fMRI, the PPM is an image of
the marginal posterior probabilities of positive activation, $\mathrm{Pr}(\beta_{j}>0\vert\mathbf{y},\mathbf{X},\mathbf{Z})$,
if the predicted BOLD is the jth covariate in $\mathbf{x}$. The joint
posterior $p(\beta,\alpha\vert\mathbf{y},\mathbf{X},\mathbf{Z})$
for the Rician and NC-$\chi$ regression models is intractable, and
we instead simulate from the joint posterior using an efficient MCMC
algorithm described in Section \ref{subsec:MCMC}. 

\subsection{Prior distribution\label{subsec:Prior}}

Our prior distribution for the Rician and the NC-$\chi$ model is
from the general class in \citet{Villani2012GSM}. Let us for clarity
focus on the prior for $\beta_{0}$ and $\beta$ in $\ln\mu_{i}=\beta_{0}+\mathbf{x}_{i}^{T}\beta$;
the prior on $\alpha_{0}$ and $\alpha$ in $\phi$ is completely
analogous. We first discuss the prior on the intercept $\beta_{0}$.
Start by standardizing the covariates to have mean zero and unit standard
deviation. This makes it reasonable to assume prior independence between
$\beta_{0}$ and $\beta$. The intercept is then $\ln\mu$ at the
mean of the original covariates. The idea is to let the user specify
a prior directly on $\mu$ when the covariates are at their means,
and then back out the implied prior on $\beta_{0}$. Let $\mu$ have
a log-normal density with mean $m^{*}$ and variance $s^{*2}$. The
induced prior on the intercept is then $\beta_{0}\sim N(m,s^{2})$
with $s^{2}=\log\left[\left(\frac{s^{*}}{m^{*}}\right){}^{2}+1\right]$
and $m=\log(m^{*})-s^{2}/2$.

The prior on $\beta$ needs some care, since its effect on the response
comes through a link function, and $\mu$ enters the model partly
via a non-linear Bessel function. Following \citet{Villani2012GSM},
we let $\beta\sim N(0,c\Sigma)$, where $\Sigma=(X^{T}\hat{D}X)^{-1}$
is the Fisher information for $\beta$, $X$ is the matrix of covariates
excluding the intercept, and $\hat{D}$ is the Fisher information
for $\mu$ conditional on $\phi$, evaluated at the prior modes of
$\beta_{0}$ and $\beta$, i.e. at the vector $(m,\boldsymbol{0}^{'})^{'}$.
Thus $\hat{D}$ depends only on the constant $m$. The conditional
Fisher information for $\mu=(\mu_{1},\dots\mu_{n})^{'}$ is a diagonal
matrix with elements 
\[
-E\left[\frac{\partial^{2}\log p(y_{i}|\mu_{i},\phi_{i})}{\partial\mu_{i}^{2}}\right]g^{\prime}(\mu_{i})^{-2}.
\]
Setting $c=n$ gives a unit information prior, i.e.\ a weak prior
that carries the information equivalent to a single observation from
the model.

\subsection{Variable selection}

Our MCMC algorithm can perform Bayesian variable selection among both
sets of covariates (i.e. $\mathbf{x}$ and $\mathbf{z}$). We make
the assumption that the intercepts in $\ln\mu$ and $\ln\phi$ are
always included in the model. Let us again focus on $\beta$ in the
equation for $\mu$. Define the vector with binary indicators $\mathcal{I}=\{I_{1},\dots I_{p}\}$
such that $I_{j}=0$ means that the $j$th element in $\beta$ is
zero, and that the corresponding covariate drops out of the model.
Let $\mathcal{I}^{c}$ denote the complement of $\mathcal{I}$. Let
$\beta_{\mathcal{I}}$ denote the subset of regression coefficients
selected by $\mathcal{I}$. To allow for variable selection we take
the previous prior $\beta_{\mathcal{}}\sim N(0,c\Sigma)$ and condition
on the zeros in $\beta$ dictated by $\mathcal{I}$: 
\begin{eqnarray*}
\beta_{\mathcal{I}}|\mathcal{I} & \sim & N\left[0,c(\Sigma_{\mathcal{I},\mathcal{I}}-\Sigma_{\mathcal{I},\mathcal{I}^{c}}\Sigma_{\mathcal{I}^{c},\mathcal{I}^{c}}^{-1}\Sigma_{\mathcal{I}^{c},\mathcal{I}}^{T})\right],
\end{eqnarray*}
and $\beta_{\mathcal{I}^{c}}|\mathcal{I}$ is identically zero. To
complete the variable selection prior, we let the elements of $\mathcal{I}$
to be a priori independent and Bernoulli distributed, i.e.$\mathrm{Pr}(I_{i}=1)=\pi$,
and $\pi$ is allowed to be different for the covariates in $\mu$
and $\phi$. We choose $\pi=0.5$ for both sets of covariates in $\mu$
and $\phi$. Other priors on $\mathcal{I}$ are just as easily handled.

\subsection{Markov Chain Monte Carlo algorithm\label{subsec:MCMC}}

We use the Metropolis-within-Gibbs sampler presented in \citet{villani2009sagm}
and \citet{Villani2012GSM}. The algorithm samples iteratively from
the set of full conditional posteriors, which in our case here are
\begin{enumerate}
\item $(\beta,\mathcal{I}_{\beta})\vert\cdot$
\item $(\alpha,\mathcal{I}_{\alpha})\vert\cdot$.
\end{enumerate}
Note that we sample $\beta$ and $\mathcal{I}_{\beta}$ jointly given
the other parameters (indicated by $\cdot$). The full conditional
posteriors $p(\beta,\mathcal{I}_{\beta}|\cdot)$ and $p(\alpha,\mathcal{I}_{\alpha}\vert\cdot)$
are highly non-standard distributions, but can be efficiently sampled
using tailored Metropolis-Hastings (MH) updates. The sampling of the
pair $(\alpha,\mathcal{I}_{\alpha})$ is analoguous to the sampling
of $(\beta,\mathcal{I}_{\beta})$, so we will only describe the update
of $(\beta,\mathcal{I}_{\beta})$. The MH proposal distribution is
of the form 
\begin{eqnarray}
J(\beta_{p},\mathcal{I}_{p}|\beta_{c},\mathcal{I}_{c}) & = & J_{1}(\beta_{p}|\mathcal{I}_{p},\beta_{c})J_{2}(\mathcal{I}_{p}|\beta_{c},\mathcal{I}_{c}),
\end{eqnarray}
where $(\beta_{c},\mathcal{I}_{c})$ denotes the current and $(\beta_{p},\mathcal{I}_{p})$
the proposed posterior draw. Following \citet{villani2009sagm} ,
we choose $J_{2}$ to be a simple proposal of $\mathcal{I}$ where
a subset of the indicators is randomly selected and a change of the
selected indicators is proposed, one variable at a time. The proposal
of $\beta$, the $J_{1}$ distribution, is a multivariate-$t$ distribution
with $\nu$ degrees of freedom: 
\begin{eqnarray*}
\beta_{p}|\mathcal{I}_{p},\beta_{c} & \sim & t_{\nu}\left[\hat{\beta},-\left(\frac{\partial^{2}\log p(\beta|\mathbf{y})}{\partial\beta\partial\beta^{T}}\right)^{-1}\bigg|_{\beta=\hat{\beta}}\right],
\end{eqnarray*}
where $\hat{\beta}$ is the terminal point of a small number of Newton
iterations to climb towards the mode of the full conditional $p(\beta_{p}|\mathcal{I}_{p},\cdot)$,
and $-\left(\frac{\partial^{2}\log p(\beta|\mathbf{y})}{\partial\beta\partial\beta^{T}}\right)^{-1}\bigg|_{\beta=\hat{\beta}}$
is the negative inverse Hessian of the full conditional posterior
evaluated at $\beta=\hat{\beta}$. Note that we are for notational
simplicity suppressing the conditioning on the covariates $\mathbf{X}$
and $\mathbf{Z}$.

There are a number of different aspects of these Newton-based proposals.
First, the number of Newton iterations can be kept very small (one
or two steps is often sufficient), since the iterations always start
at $\beta_{c}$, which is typically not far from the mode. Second,
$\hat{\beta}$ is often not exactly the mode, but the posterior draws
from the algorithm will nevertheless converge to the underlying target
posterior. Third, the update $(\beta_{c},\mathcal{I}_{c})\rightarrow(\beta_{p},\mathcal{I}_{p})$
is accepted with probability
\[
\min\left(1,\frac{p(y|\beta_{p},\mathcal{I}_{p})p(\beta_{p}|\mathcal{I}_{p})p(\mathcal{I}_{p})/J_{1}(\beta_{p}|\mathcal{I}_{p},\beta_{c})J_{2}(\mathcal{I}_{p}|\beta_{c},\mathcal{I}_{c})}{p(y|\beta_{c},\mathcal{I}_{c})p(\beta_{c}|\mathcal{I}_{c})p(\mathcal{I}_{c})/J_{1}(\beta_{c}|\mathcal{I}_{c},\beta_{p})J_{2}(\mathcal{I}_{c}|\beta_{p},\mathcal{I}_{p})}\right),
\]
where the factor $J_{1}(\beta_{c}|\mathcal{I}_{c},\beta_{p})$ is
computed from another round of Newton iterations, this time starting
from the proposed point $\beta_{p}$. Fourth, to implement the Newton
iterations we need to be able to compute the gradient $\frac{\partial\log p(y|\beta)}{\partial\beta}$
and the Hessian $\frac{\partial^{2}\log p(\beta|y)}{\partial\beta\partial\beta^{T}}$
efficiently. \citet{Villani2012GSM} show that this can be done very
efficiently using the chain rule and compact matrix computations,
and Appendix A gives the details for the NC-$\chi$ regression. In
DTI, when the parameter space is restricted to the set of positive
definite matrices, these expressions need to be extended, see Section
\ref{sec:Tensor}.

In summary, our proposed algorithm consists of a two-block Metropolis-Hastings
within Gibbs sampler, where each updating step updates a set of regression
coefficients simultaneously with their binary variable selection indicators.
The multivariate student-$t$ proposal is tailored to the full conditional
posterior at each step, using a Newton method to approximate the conditional
posterior mode and curvature (Hessian). The computations are very
fast since the gradient and the Hessian for the Newton steps can be
computed very efficiently in compact matrix form, and only a very
small number of Newton steps is needed, since each iteration starts
at the previously accepted parameter draw which is typically an excellent
initial value.

\section{Activity localization in fmri data\label{sec:ActivityfMRI}}

\begin{figure}
\begin{centering}
\includegraphics[scale=0.5]{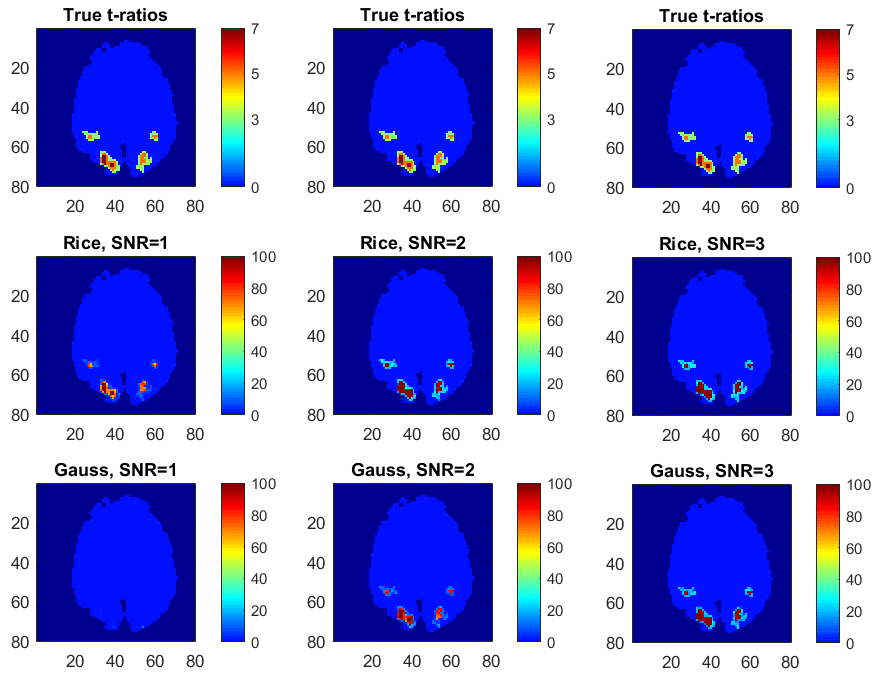}
\par\end{centering}
\caption{Comparison of activation inferences using the Rician and Gaussian
models on simulated fMRI data. Top row: True activations in the Rician
data generating model. Middle (Rice) and bottom (Gauss) row: Percentage
of simulated datasets where the posterior probability of activation
is larger than 99 \%. \label{Activity99}}
\end{figure}

\begin{figure}
\begin{centering}
\includegraphics[scale=0.5]{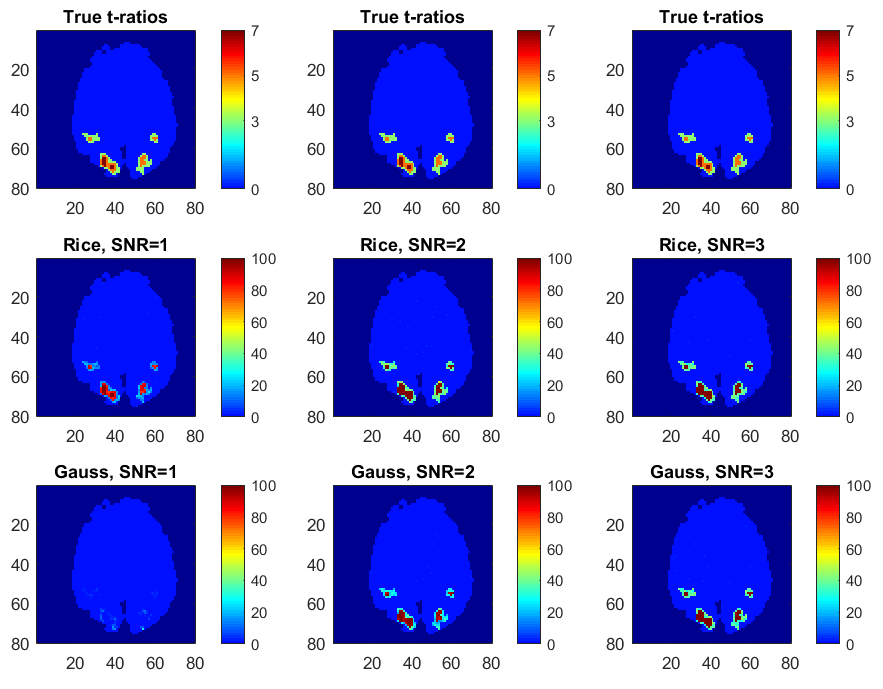}
\par\end{centering}
\caption{Comparison of activation inferences in the Rician and Gaussian models
using simulated fMRI data. Top row: True activations in the Rician
data generating model. Middle (Rice) and bottom (Gauss) row: Percentage
of simulated datasets where the posterior probability of activation
is larger than 95 \%. \label{Activity95}}
\end{figure}

Comparisons of the proposed Rician model (Eq. \ref{eq:RiceModel})
to a corresponding Gaussian model using several commonly used fMRI
datasets showed no detectable differences between the two models since
the SNRs were larger than three in all voxels; this is in line with
the results in \citep{RoweLogan2004}. As discussed in the Introduction,
however, there are situations when SNRs can be low in fMRI, in particular
for high-resolution imaging. We therefore compare the two models using
simulated fMRI data with Rician noise at the three different SNR levels
(1, 2 and 3). The data are simulated from a model that mimic the results
from a real fMRI experiment with a simple block paradigm. The real
fMRI data had a spatial resolution of 1.6 x 1.6 x 1.8 mm$^3$, and
the noise variance and the variance for the activation parameter was
manipulated in the simulated datasets to obtain a pre-specified level
of activation and SNR. The noise variance experimentally controls
the SNR levels, while the variance for the activation parameter is
adjusted to accommodate one of the four chosen $t$-ratios (0, 3,
5 and 7) for each of the voxels on our selected slice of the brain.
The prior distributions on the parameters in the Rician and Gaussian
model are carefully chosen to carry the same information in both models.
Specifically, we choose unit information priors (see Section 3.2),
such that the priors only carry the information from a single observation
in each of the models. We simulate 100 datasets for each SNR level
(1, 2, 3). The first row with graphs in Figures \ref{Activity99}
and \ref{Activity95} shows the four activated regions in the data
generating process in the form of ``$t$-ratios'' (parameter value/standard
deviation). The second row of graphs show that all four activation
regions are correctly localized with our Rician model in a large majority
of the simulated datasets, while the third row shows that the Gaussian
model completely misses all of the activated regions when SNR=1, and
has a high failure rate when SNR=2.

\clearpage

\section{Estimating fractional anisotropy and mean diffusivity in DWI data}

\subsection{Diffusion weighted imaging\label{sec:DWI}}

While fMRI data are mainly specified by the echo time and the repetition
time of the pulse sequence, DWI data also require specification of
the $b$-value \citep{bihan2}. The $b$-value in turn depends on
two factors; the strength and the duration of the diffusion gradient.
Using a larger $b$-value enables more advanced diffusion models,
e.g. through HARDI \citep{tuch}, which for example can be used to
properly account for multiple fiber orientations in a single voxel.
A significant drawback of a higher $b$-value is, however, a lower
signal to noise ratio. The main reason for this is that the signal
decays exponentially with time, and high b-values require longer diffusion
gradients. As a consequence, Rician noise models are far more common
for DWI than for fMRI, as the Rician distribution is only well approximated
by a Gaussian for high SNRs.

\subsection{The diffusion tensor model\label{sec:Tensor}}

The most common diffusion tensor model states that the signal $S_{i}$
for measurement $i$ can be written as
\begin{equation}
S_{i}=S_{0}\exp\left(-b_{i}g_{i}^{T}\mathbf{D}g_{i}\right),\label{eq:DiffTensorModel}
\end{equation}
where $S_{0}$ is the signal in absence of any diffusion gradient,
$b_{i}$ is the $b$-value, $g_{i}=(g_{ix},g_{iy},g_{iz})^{T}$ is
the gradient vector and 
\[
\mathbf{D}=\left(\begin{array}{ccc}
d_{xx} & d_{xy} & d_{xz}\\
d_{xy} & d_{yy} & d_{yz}\\
d_{xy} & d_{yz} & d_{zz}
\end{array}\right)
\]
is the diffusion tensor. The single-diffusion tensor model in \eqref{eq:DiffTensorModel}
can be written as a regression model of the form in Eq. \eqref{eq:NC-model}
with (see e.g. \citet{KoayBookChapter})
\begin{equation}
\ln\mu_{i}=\beta_{0}+\mathbf{x}_{i}^{T}\beta,\label{eq:logMuRegressionDWI}
\end{equation}
where $\beta_{0}=\ln S_{0}$, $\beta=\left(d_{xx},d_{yy},d_{zz},d_{xy},d_{yz},d_{xz}\right)$
and 
\[
\mathbf{x}_{i}^{T}=-\left(b_{i}g_{ix}^{2},b_{i}g_{iy}^{2},b_{i}g_{iz}^{2},2b_{i}g_{ix}g_{iy},2b_{i}g_{iy}g_{iz},2b_{i}g_{ix}g_{iz}\right).
\]

For single-coil imaging, the noise around $\mu_{i}$ is Rician, and
cannot be well approximated by a Gaussian model for high $b$-values
where the signal-to-noise ratio is low. When data are collected by
parallel techniques using $L$ coils, the noise is either Rician distributed
or NC-$\chi$ distributed with $L$ degrees of freedom. If the composite
signal is a complex weighted sum of the $L$ signals, the magnitude
of the composite signal is Rician distributed. If the simpler sum
of squares approach is used for merging the L signals into a single
image, the resulting signal is instead NC-$\chi$ distributed \citep{tristan2012least,fernandez2016}.

Note that since the tensor $D$ is positive definite, the parameter
space of $\beta$ in \eqref{eq:logMuRegressionDWI} is restricted.
One can impose the positive definitness restriction by assigning zero
prior probability to all $\beta$ that correspond to a negative definite
$D$; all such proposals will then be rejected in the MCMC. This may,
however, lead to excessive rejections, and a better solution is to
impose the positive definiteness restriction explicitly. We here use
the Log-Cholesky representation \citep{KoayBookChapter}, where the
diffusion tensor $D$ is expressed as 
\[
D(\omega)=\Omega^{T}\Omega
\]
with 
\[
\Omega=\left(\begin{array}{ccc}
e^{\omega_{1}} & \omega_{4} & \omega_{6}\\
0 & e^{\omega_{2}} & \omega_{5}\\
0 & 0 & e^{\omega_{3}}
\end{array}\right).
\]
In this parametrization the tensor can be written as
\[
D(\omega)=\left(\begin{array}{ccc}
e^{2\omega_{1}} & \omega_{4}e^{\omega_{1}} & \omega_{6}e^{\omega_{1}}\\
\omega_{4}e^{\omega_{1}} & \omega_{4}^{2}+e^{2\omega_{2}} & \omega_{4}\omega_{6}+\omega_{5}e^{\omega_{2}}\\
\omega_{6}e^{\omega_{1}} & \omega_{4}\omega_{6}+\omega_{5}e^{\omega_{2}} & \omega_{6}^{2}+\omega_{5}^{2}+e^{2\omega_{3}}
\end{array}\right),
\]
such that the vector of regression coefficients $\beta(\omega)$ is
given by
\[
(e^{2\omega_{1}},\omega_{4}^{2}+e^{2\omega_{2}},\omega_{6}^{2}+\omega_{5}^{2}+e^{2\omega_{3}},\omega_{4}e^{\omega_{1}},\omega_{4}\omega_{6}+\omega_{5}e^{\omega_{2}},\omega_{6}e^{\omega_{1}}).
\]

Most applications with the diffusion tensor model takes the logarithm
of the measurements and estimates $\beta$ with least squares (see
\citet{KoayBookChapter} for an overview). This estimation method
therefore does not respect the log link in the mean. One can also
argue that it also implicitly assumes Gaussian noise in the sense
that least squares equals the maximum likelihood estimate only when
the noise is Gaussian. Moreover, it does not guarantee that the estimated
tensor is positive definite. We refer to \citet{KoayBookChapter}
for an overview of constrained non-linear least squares alternatives. 

We will here take a Bayesian approach with Rician or NC-$\chi$ noise,
using a proper log link and a parametrization that guarantees that
the posterior mass is fully contained within the space of positive
definite matrices. Existing Bayesian approaches to DTI assume Gaussian
noise and use the random walk Metropolis (RWM) algorithm to simulate
from the posterior distribution. RWM is easy to implement, but is
well known to explore the posterior distribution very slowly (see
Section \ref{sec:ResultsDTI}). The Metropolis-within-Gibbs algorithm
with tailored proposals and variable selection to reduce the dimensionality
of the parameter space presented in Section \ref{subsec:MCMC} can
explore the posterior distribution in a much more efficient manner
\citep{villani2009sagm,Villani2012GSM}. As a result of the non-linear
mapping from $\omega$ to $\beta$, the gradient of the likelihood
part of Equation \ref{General Gradient} is modified to 
\[
\frac{\partial\ln p(y|\omega)}{\partial\omega}=\left(\mathbf{X}\frac{\partial\beta(\omega)}{\partial\omega}\right)^{T}\mathbf{g},
\]
where 
\[
\frac{\partial\beta(\omega)}{\partial\omega}=\left(\begin{array}{cccccc}
2e^{2\omega_{1}} & 0 & 0 & 0 & 0 & 0\\
0 & 2e^{2\omega_{2}} & 0 & 2\omega_{4} & 0 & 0\\
0 & 0 & 2e^{2\omega_{3}} & 0 & 2\omega_{5} & 2\omega_{6}\\
\omega_{4}e^{\omega_{1}} & 0 & 0 & e^{\omega_{1}} & 0 & 0\\
0 & \omega_{5}e^{\omega_{2}} & 0 & \omega_{6} & e^{\omega_{2}} & \omega_{4}\\
\omega_{6}e^{\omega_{1}} & 0 & 0 & 0 & 0 & e^{\omega_{1}}
\end{array}\right).
\]
The Hessian in Equation \ref{General Hessian} can be modified accordingly. 

The Fisher information based prior presented in Section \ref{subsec:Prior}
can in principle be used for DTI. We have found however that the numerical
stability of our MCMC sampler improves if we use an alternative prior,
which we now describe. We assume the priors for the intercepts $\beta_{0}\sim N(m_{\beta},d)$
and $\alpha_{0}\sim N(m_{\alpha},d)$, independently of the priors
for the unrestricted tensor coefficients $\omega\sim N(0,cI)$ and
the parameters variance function $\alpha\sim N(0,cI)$, where $c=100$
to induce non-informative priors and $I$ is the identity matrix.
Note that the prior expected value of $0$ for $\alpha$ implies that
the variance of the underlying complex-valued signal $\phi$ is centered
on the homoscedastic model a priori. To set the prior mean on the
intercepts $\beta_{0}$ and $\alpha_{0}$, note first that the models
for $\mu$ and $\sigma^{2}$ in Eq. \ref{eq:RiceModel} become $\beta_{0}=\ln\mu_{i}$
and $\alpha_{0}=\ln\sigma_{i}$ when $b=0$. It is therefore common
in the DTI literature to separately pre-estimate the mean intercept
$\beta_{0}$ by the logarithm of the mean of measurements $y$ when
$b=0$, and then subsequently remove these observations from the dataset.
This procedure improves the numerical stability of the estimations.
In a similar vein, we set the prior expected values, $m_{\beta}$
and $m_{\alpha}$ by taking the logarithm of the mean and variance
of $y$ when $b=0,$ respectively; the observations with zero $b$-values
are then removed from the dataset in the remaining estimation. We
have found improved numerical stability in the MCMC algorithm if we
allow for a positive, but small, prior variance of $d=0.1$.

\subsection{Data\label{sec:DTIData}}

We use the freely available MGH adult diffusion dataset from the Human
Connectome Project (HCP) \citep{Setsompop,VanEssen} \footnote{http://www.humanconnectome.org/documentation/MGH-diffusion/}.
The dataset comprise DWI data collected with several different $b$-values,
and the downloaded data have already been corrected for gradient nonlinearities,
subject motion and eddy currents \citep{Glasser,Andersson2016}. The
DWI data were collected using a spin-echo EPI sequence and a 64-channel
array coil \citep{Setsompop}, yielding volumes of 140 x 140 x 96
voxels with an isotropic voxel size of 1.5 mm. The data collection
was divided into 5 runs, giving data with four different $b$-values:
1,000, 3,000, 5,000, and 10,000 s/mm$^{2}$. The number of gradient
directions was 64 for $b$ = 1,000 s/mm$^{2}$ and $b$ = 3,000 s/mm$^{2}$,
128 for $b$ = 5,000 s/mm$^{2}$, and 256 for $b$ = 10,000 s/mm$^{2}$.
Merging the measurements from the 64 channels into a single image
was performed using a complex weighted combination \citep{Setsompop},
instead of the more simple sum of squares approach. This is an important
fact, as the weighted approach for this data leads to noise with a
Rician distribution, instead of the NC-$\chi$ distribution resulting
from the sum of squares approach \citep{fernandez2016}. Prior to
any statistical analysis, the function FAST \citep{Zhang} in FSL
was used to generate a mask of white brain matter, gray brain matter
and cerebrospinal fluid (CSF), to avoid running the analysis on voxels
in CSF.

Data used in the preparation of this work were obtained from the Human
Connectome Project (HCP) database (https://ida.loni.usc.edu/login.jsp).
The HCP project (Principal Investigators: Bruce Rosen, M.D., Ph.D.,
Martinos Center at Massachusetts General Hospital; Arthur W. Toga,
Ph.D., University of Southern California, Van J. Weeden, MD, Martinos
Center at Massachusetts General Hospital) is supported by the National
Institute of Dental and Craniofacial Research (NIDCR), the National
Institute of Mental Health (NIMH) and the National Institute of Neurological
Disorders and Stroke (NINDS). HCP is the result of efforts of co-investigators
from the University of Southern California, Martinos Center for Biomedical
Imaging at Massachusetts General Hospital (MGH), Washington University,
and the University of Minnesota.

\subsection{Comparisons between the Rician and Gaussian DTI models\label{sec:ResultsDTI}}

We compare the Rician and Gaussian DTI models for the voxels in slice
$50$ in the middle of the brain. We mainly compare the estimation
results between the models using the whole dataset with all b-values
up to $b$ = 10,000 s/mm$^{2}$, but also show some results for subsets
of the whole dataset with b-values up to $b$ = 3,000 s/mm$^{2}$
and $b$ = 5,000 s/mm$^{2}$, respectively. The expected Hessian is
used for the MH proposals of the parameters in the Gaussian case,
but since the expected Hessian is not available for the Rician model,
different combinations of the observed Hessian and the outer product
of gradients for $\mu$ and $\phi$ are used in each voxel, to improve
the numerical stability of the estimations. Our MCMC convergence is
excellent for both the Rician and Gaussian DTI models, with high acceptance
probabilities for $\mu$ and $\phi$ in almost all voxels for all
estimated datasets. The mean MH acceptance probabilities for $\mu$
and $\phi$ are 74 \% and 87 \% for the Rician model, compared to
70 \% and 90 \% for the Gaussian model. The standard deviations of
the acceptance probabilities across voxels are 7.5 \% and 16.2 \%
for the Rician model, compared to 5.1 \% and 5.2 \% for the Gaussian
model.

\begin{figure}
\begin{centering}
\includegraphics[scale=0.4]{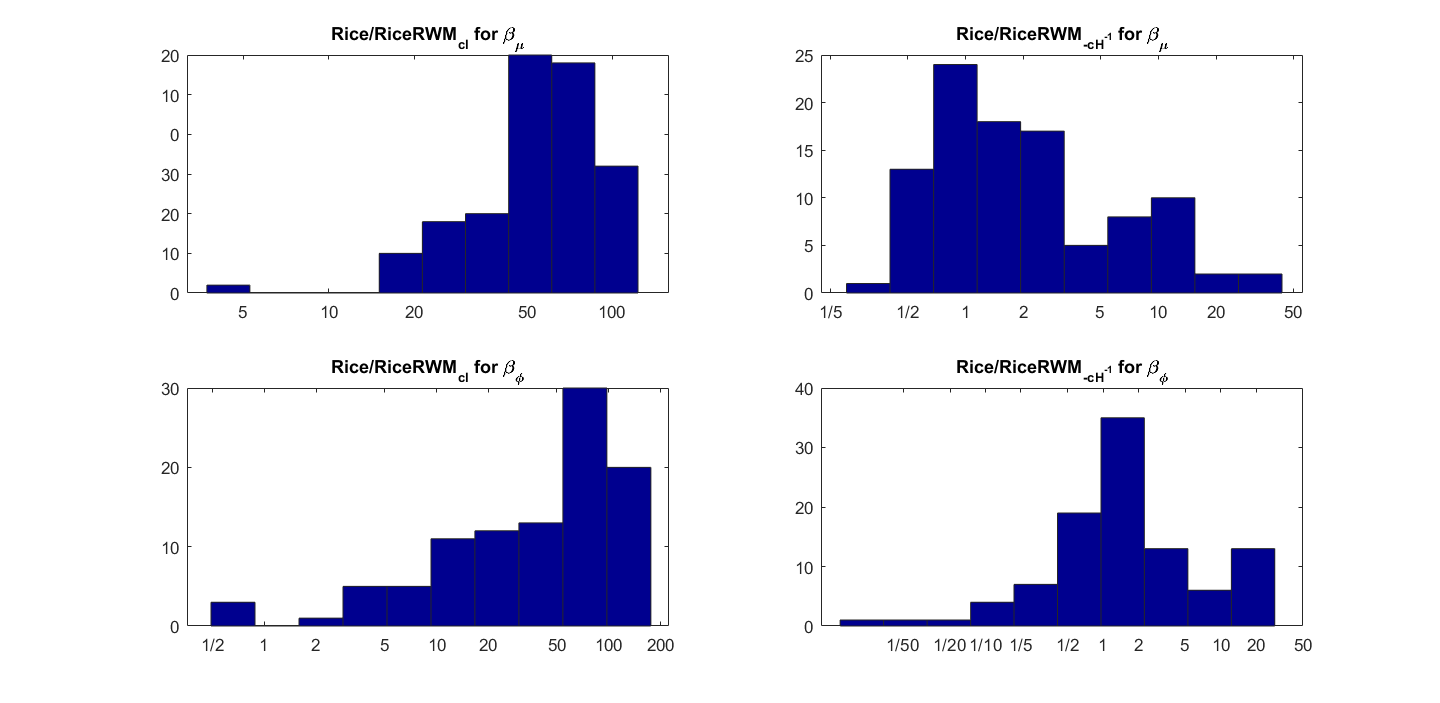}
\par\end{centering}
\caption{Histograms of the ratio of independent draws per minute for our MCMC
algorithm compared to each type of RWM algorithm for 100 randomly
sampled white matter voxels for the Rician model. The rows correspond
to the parameters, the columns to the two covariance matrices in the
RWM algorithm. \label{fig:IndepDrawsRice}}
\end{figure}

\begin{figure}
\begin{centering}
\includegraphics[scale=0.4]{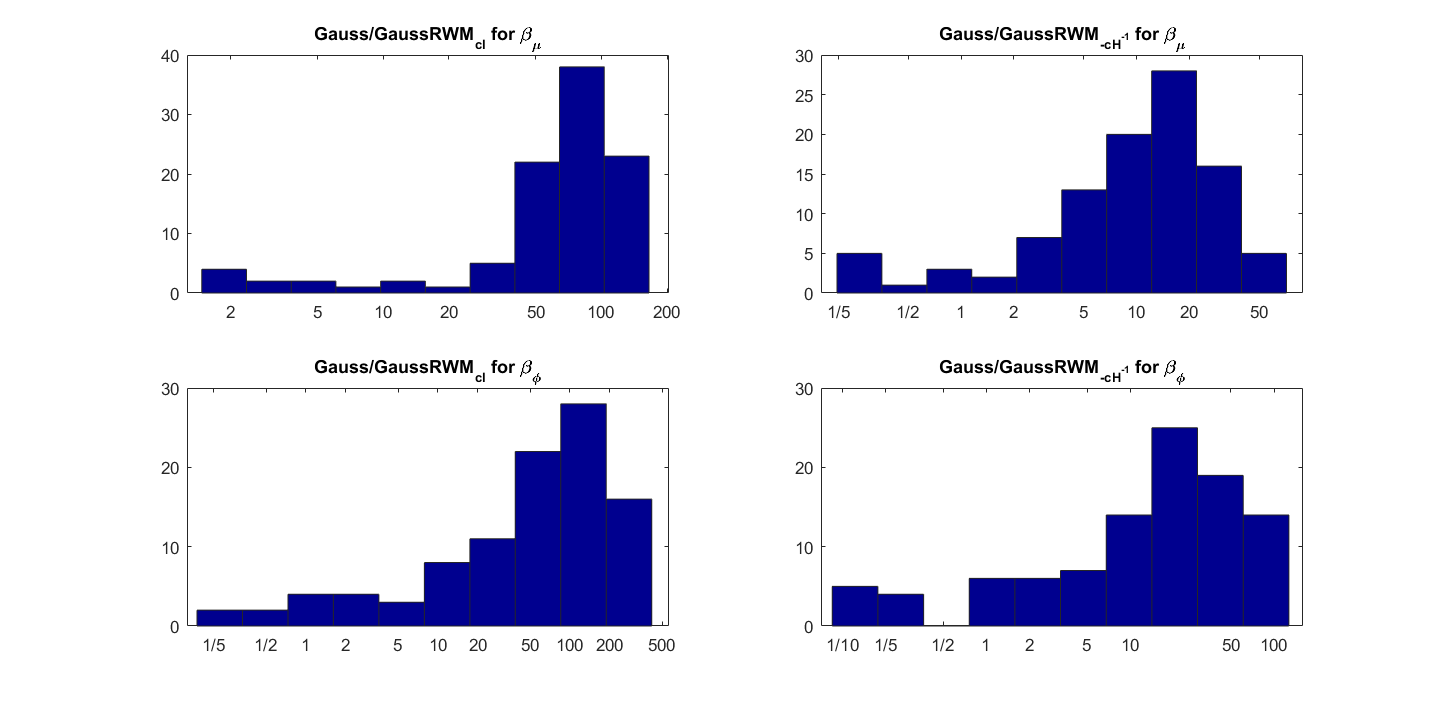}
\par\end{centering}
\caption{Histograms of the ratio of independent draws per minute for our MCMC
algorithm compared to each type of RWM algorithm for 100 randomly
sampled white matter voxels for the Gaussian model. The rows correspond
to the parameters, the columns to the two covariance matrices in the
RWM algorithm. \label{fig:IndepDrawsGauss}}
\end{figure}

We compare the efficiency of our MCMC algorithm to commonly used Random
Walk Metropolis (RWM) algorithms for MCMC in DTI (see e.g. the highly
influential work by \citet{behrens2003}). The RWM algorithms use
a multivariate normal distribution centered on the current parameter
value to propose a posterior draw of all parameters in $\mu$ and
$\sigma$ in a single block. The most common choice of proposal covariance
matrix in DTI is a scaled identity matrix where the scale is chosen
adaptively to achieve optimal performance. We also compare our MCMC
algorithm to a refined version with covariance matrix $-c$$H^{-1}$,
where $H$ is the Hessian at the posterior mode and $c$ is a scalar
which is again chosen adaptively for optimal performance. Using the
posterior results from 100 randomly sampled white matter voxels, Figures
\ref{fig:IndepDrawsRice} and \ref{fig:IndepDrawsGauss} show histograms
of the ratios of the number of independent draws per minute for our
MCMC algorithm compared to each type of RWM algorithm. Results are
presented for both the Rician and Gaussian models. The number of independent
MCMC draws is defined as the number of total MCMC draws divided by
the estimated inefficiency factor $IF=1+2\sum_{k=1}^{\infty}\rho_{k},$
where $\rho_{k}$ is the autocorrelation function of lag $k$ of the
MCMC chain.

In general, our MCMC algorithm is much more efficient in almost all
voxels than the RWM algorithm with covariance matrix $cI$ for both
the Rician and Gaussian models. In most voxels, our MCMC algorithm
for the Gaussian model is also more efficient than the RWM algorithm
with covariance matrix $-cH^{-1}$. This is also true for our MCMC
algorithm in a majority of the voxels for the Rician model, especially
for parameters $\beta_{\mu}$ compared to $\beta_{\sigma}$. In a
random sample of 100 gray matter voxels, we also find that our MCMC
algorithm is much more efficient than the RWM algorithm with covariance
matrix $cI$, but compared to the RWM algorithm with covariance matrix
$-cH^{-1}$ our algorithm is only slightly better in both models (not
shown here). 

\begin{figure}
\begin{centering}
\includegraphics[scale=0.42]{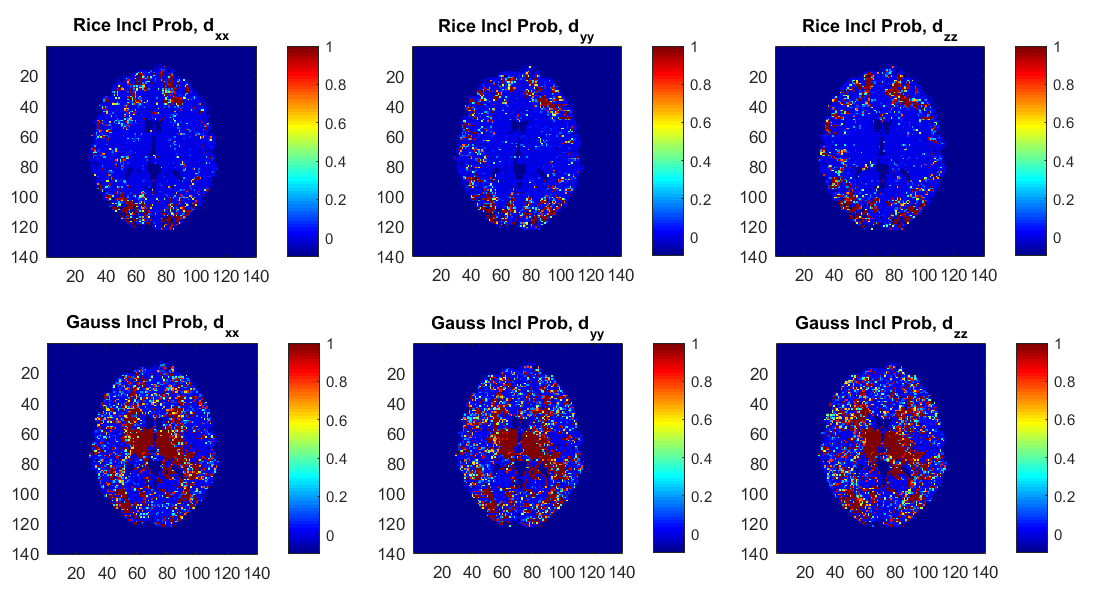}
\par\end{centering}
\caption{Posterior inclusion probabilities for the covariates corresponding
to the diffusion directions $\left(d_{xx},d_{yy},d_{zz}\right)$ in
the variance function $\phi$ for the Rician and Gaussian DTI models.
Note that the inclusion probability is close to 1 for a large number
of voxels, especially for the Gaussian model. \label{fig:PostInclProbs}}
\end{figure}

Figure \ref{fig:PostInclProbs} shows posterior inclusion probabilities
for the covariates corresponding to \\
$\left(d_{xx},d_{yy},d_{zz}\right)$ in $\mathbf{z}$ (the variance
function) for both models. In a large number of voxels the inclusion
probabilities for the Gaussian model are close or equal to $1$, compared
to far fewer voxels for the Rician model. In addition, there are substantially
more voxels with this property in the mid-regions of the brain for
the Gaussian model, and in the outer parts of the brain for the Rician
model. The inclusion probabilities for the remaining covariates in
$\mathbf{z}$ are in most cases very close to zero for both models,
especially for the Gaussian model (not shown here). This clearly shows
that diffusion covariates affect the noise variance in both models,
and may imply that homoscedastic DTI models can give distorted results
as documented in \citet{WegmannHetero} for the Gaussian DTI model.
Using a part of the dataset with all b-values up to $b$ = 5,000 s/mm$^{2}$
(thereby excluding relatively uncommon measurements at a b-value of
10,000) implies far fewer voxels with inclusion probabilities close
or equal to $1$ for both models, but there are still substantially
more voxels with this property for the Gaussian model compared to
the Rician model (not shown here).

\noindent 

The estimated single-diffusion tensors are compared across voxels
for the Rician DTI model in Eq. \ref{eq:RiceModel} to the Gaussian
counterpart, with respect to the DTI scalar measures mean diffusivity
(MD) and fractional anisotropy (FA). The DTI scalar measures are functions
of the eigenvalues $\lambda_{1}\geq\lambda_{2}\geq\lambda_{3}$ of
the single-diffusion tensor, defined as

\[
MD=\frac{\lambda_{1}+\lambda_{2}+\lambda_{3}}{3},\;FA=\sqrt{\frac{3}{2}}\sqrt{\frac{\stackrel[i=1]{3}{\sum}\left(\lambda_{i}-MD\right)^{2}}{\stackrel[i=1]{3}{\sum}\lambda_{i}^{2}}}.
\]

\begin{figure}
\begin{centering}
\includegraphics[scale=0.68]{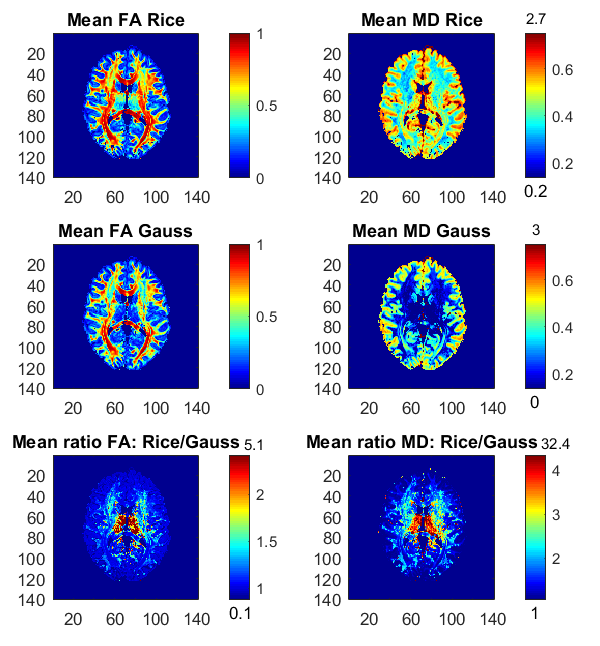}
\par\end{centering}
\caption{Posterior means and ratios of posterior means of FA and MD for the
Rician and Gaussian DTI models, using the whole dataset with all b-values
up to $b$ = 10,000 s/mm$^{2}$. The colorbars are shown for the mid
95 \% values and the minimum and maximum values are marked out at
the bottom and top of the colorbars, respectively. \label{fig:MeanFAMD10000}}
\end{figure}

\begin{figure}
\begin{centering}
\includegraphics[scale=0.68]{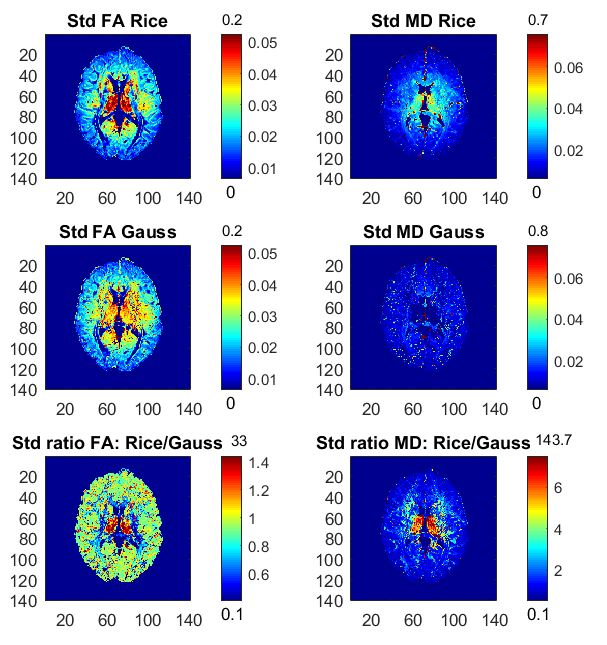}
\par\end{centering}
\caption{Posterior standard deviations and ratios of posterior standard deviations
of FA and MD for the Rician and Gaussian DTI models, using the whole
dataset with all b-values up to $b$ = 10,000 s/mm$^{2}$. The colorbars
are shown for the mid 95 \% values and the minimum and maximum values
are marked out at the bottom and top of the colorbars, respectively. \label{fig:StdDevFAMD10000}}
\end{figure}

\noindent Figure \ref{fig:MeanFAMD10000} shows the posterior means
of FA and MD and the ratios of posterior means between the models,
and Figure \ref{fig:StdDevFAMD10000} shows the posterior standard
deviations of FA and MD and the ratios of posterior standard deviations
between the models. 
 
 In general, the Gaussian model substantially underestimates mean
values of FA in many voxels, especially in mid-regions with low or
mid-size values of FA, compared to the theoretically correct Rician
model. In addition, the Gaussian model greatly underestimates MD across
the whole slice of the brain compared to the Rician model. Hence,
using the Gaussian model for DTI can therefore lead to severely misleading
inferences. The standard deviations of FA and MD are small for both
models. In white-matter regions with high FA values the Gaussian model
estimates slightly larger standard deviations of FA compared to mid-regions
with slightly larger standard deviations of FA for the Rician model.
On the other hand, the standard deviations of MD are underestimated
by the Gaussian model in all voxels.

\begin{figure}
\begin{centering}
\includegraphics[scale=0.68]{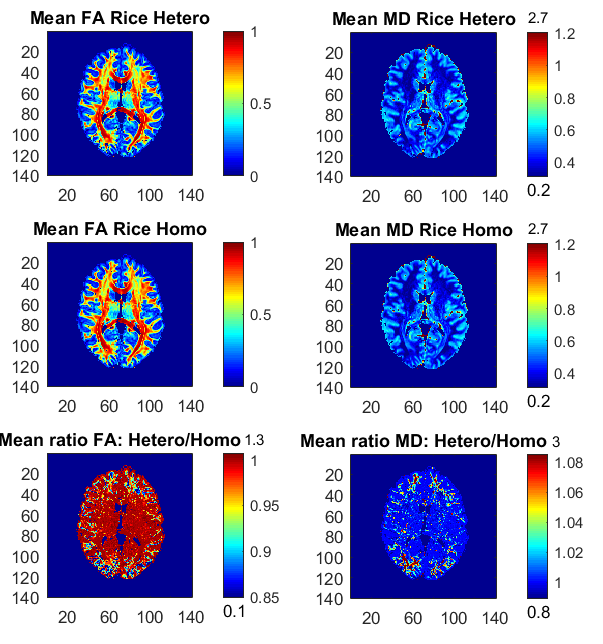}
\par\end{centering}
\caption{Posterior means and ratios of posterior means of FA and MD for the
heteroscedastic (Hetero) and homoscedastic (Homo) Rician DTI models,
using the whole dataset with all b-values up to $b$ = 10,000 s/mm$^{2}$.
The colorbars are shown for the mid 95 \% values and the minimum and
maximum values are marked out at the bottom and top of the colorbars,
respectively.  \label{fig:MeanRatioFAMDHomoHetero}}
\end{figure}

Figure \ref{fig:MeanRatioFAMDHomoHetero} shows the posterior means
of FA and MD and the ratios of posterior means for the Rician models
with covariates in the noise variance $\phi$ (heteroscedastic model)
and without covariates in $\phi$ (homoscedastic model). The differences
between the models are small, but in the outer parts of the brain
the homoscedastic Rician model slightly overestimates the posterior
means of FA in a large number of voxels. The posterior standard deviations
of FA and MD for the Rician models are similar, but the homoscedastic
Rician model slightly underestimates, in general, the standard deviation
of FA in the outer parts of the brain (not shown here). The differences
in FA between the Rician models agree with our previous findings that
the diffusion covariates (directions) especially affect the noise
variance for the Rician model in the outer parts of the brain, where
directional DTI measures such as FA are affected. This is in contrast
to the non-directional measure MD, for which the differences between
the models are negligible. Hence, in voxels with heteroscedastic noise
variance that depends on the diffusion directions the posterior means
and standard deviations of FA are slightly different for the heteroscedastic
and homoscedastic Rician models.

\noindent 

\begin{figure}
\begin{centering}
\includegraphics[scale=0.68]{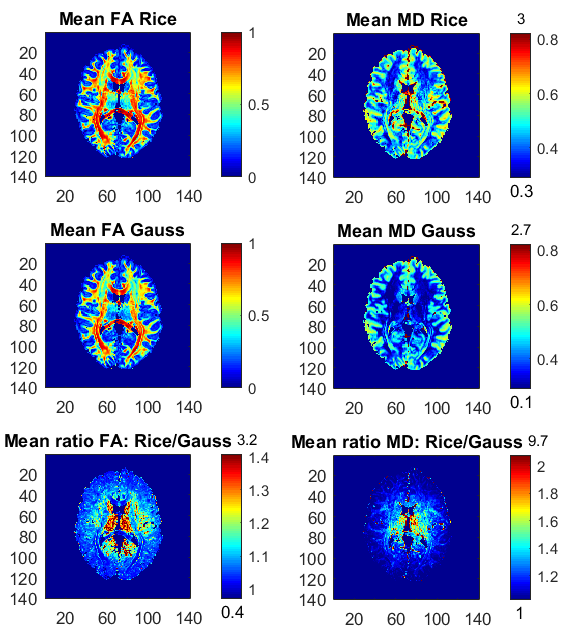}
\par\end{centering}
\caption{Posterior means and ratios of posterior means of FA and MD for the
Rician and Gaussian DTI models, using the part of the dataset with
all b-values up to $b$ = 5,000 s/mm$^{2}$. The colorbars are shown
for the mid 95 \% values and the minimum and maximum values are marked
out at the bottom and top of the colorbars, respectively.  \label{fig:MeanRatioFAMD5000}}
\end{figure}

It is relatively uncommon with measurements at a b-value of 10,000.
Figure \ref{fig:MeanRatioFAMD5000} therefore shows the posterior
means and the ratios of posterior means of FA and MD between the models
for the part of the whole dataset with all b-values up to $b$ = 5,000
s/mm$^{2}$, hence excluding the observations with the highest b-value.
The differences in FA and MD are notably smaller compared to the results
from the whole dataset, but the Gaussian model still underestimates
the posterior mean values of FA and MD substantially in many voxels.
Hence, using the Gaussian model can also lead to misleading inferences
for this smaller subset of the data. Taking an even smaller data subset
with all b-values up to $b$ = 3,000 s/mm$^{2}$, the differences
in FA and MD between the models become negligible, where the Gaussian
model only slightly underestimates FA and MD in some voxels (not shown
here).

\noindent 

\begin{figure}
\begin{centering}
\includegraphics[scale=0.4]{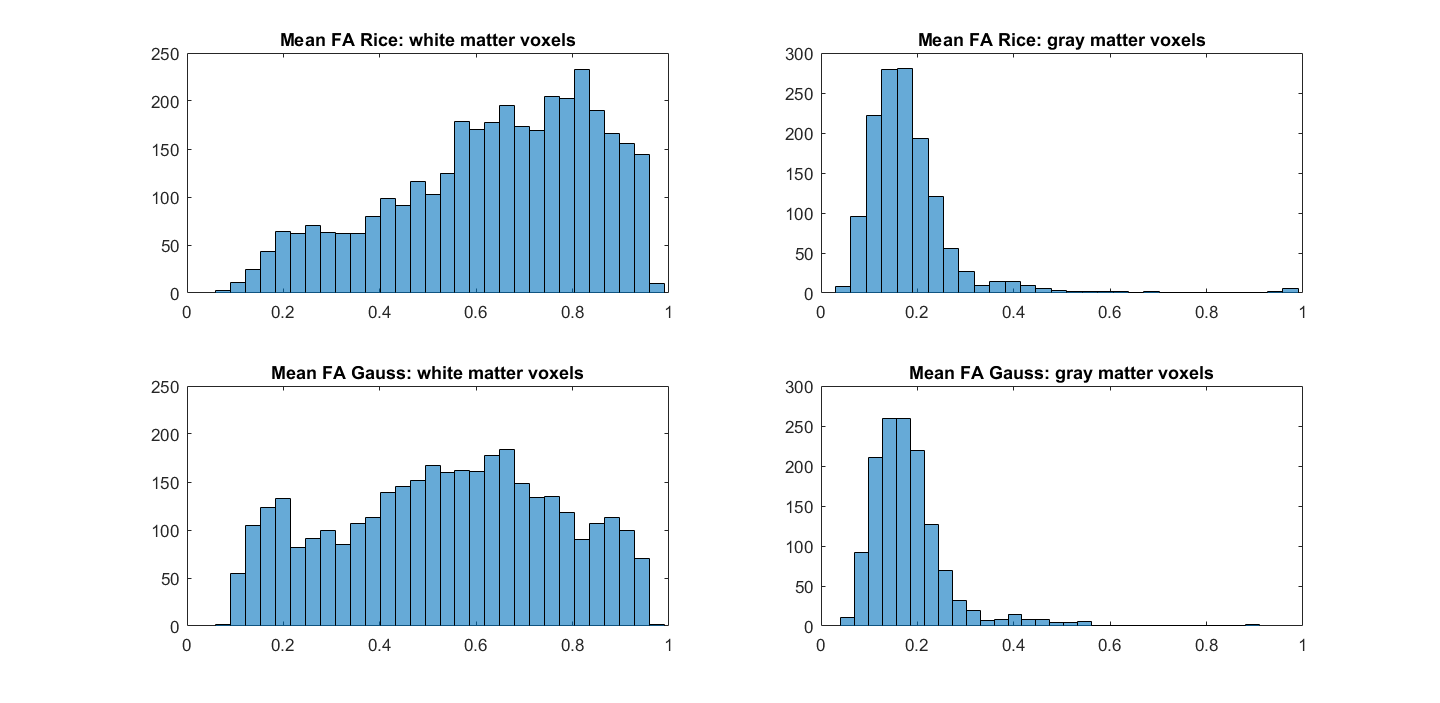}
\par\end{centering}
\caption{Histograms of posterior means of FA for white-matter (left) and gray-matter
(right) voxels for the Rician and Gaussian DTI models, using the whole
dataset. A white-matter (gray-matter) voxel is defined as a voxel
where the probability is 1 for white matter (gray matter) from the
function FAST in FSL.\label{fig:WhiteGrayMeanFA}}
\end{figure}

To investigate the differences between the Rician and Gaussian models
in white and gray matter, we use the function FAST in FSL to compute
the probabilities for white matter, gray matter and CSF in each voxel
of the brain. Let a white-matter (gray-matter) voxel be defined as
a voxel where the probability is 1 for white matter (gray matter).
It is generally expected that white-matter voxels have higher FA,
compared to gray-matter voxels. Figure \ref{fig:WhiteGrayMeanFA}
shows that this is true for the Rician model as the distribution of
the posterior means of FA is more skewed to larger values, compared
to more uniformly distributed posterior means of FA for the Gaussian
model. Hence, the Gaussian model underestimates, on average, FA in
white-matter voxels. In addition, Figure \ref{fig:WhiteGrayMeanFA}
shows that the uncertainty of FA for white-matter voxels is somewhat
lower for the Rician model compared to the Gaussian model, with the
distribution of the standard deviations of FA being more skewed to
the left for the Rician model. This is in contrast to the gray-matter
voxels, where the distributions of both the posterior means and standard
deviations of FA are very similar and concentrated at low values for
both models.

\begin{figure}
\begin{centering}
\includegraphics[scale=0.4]{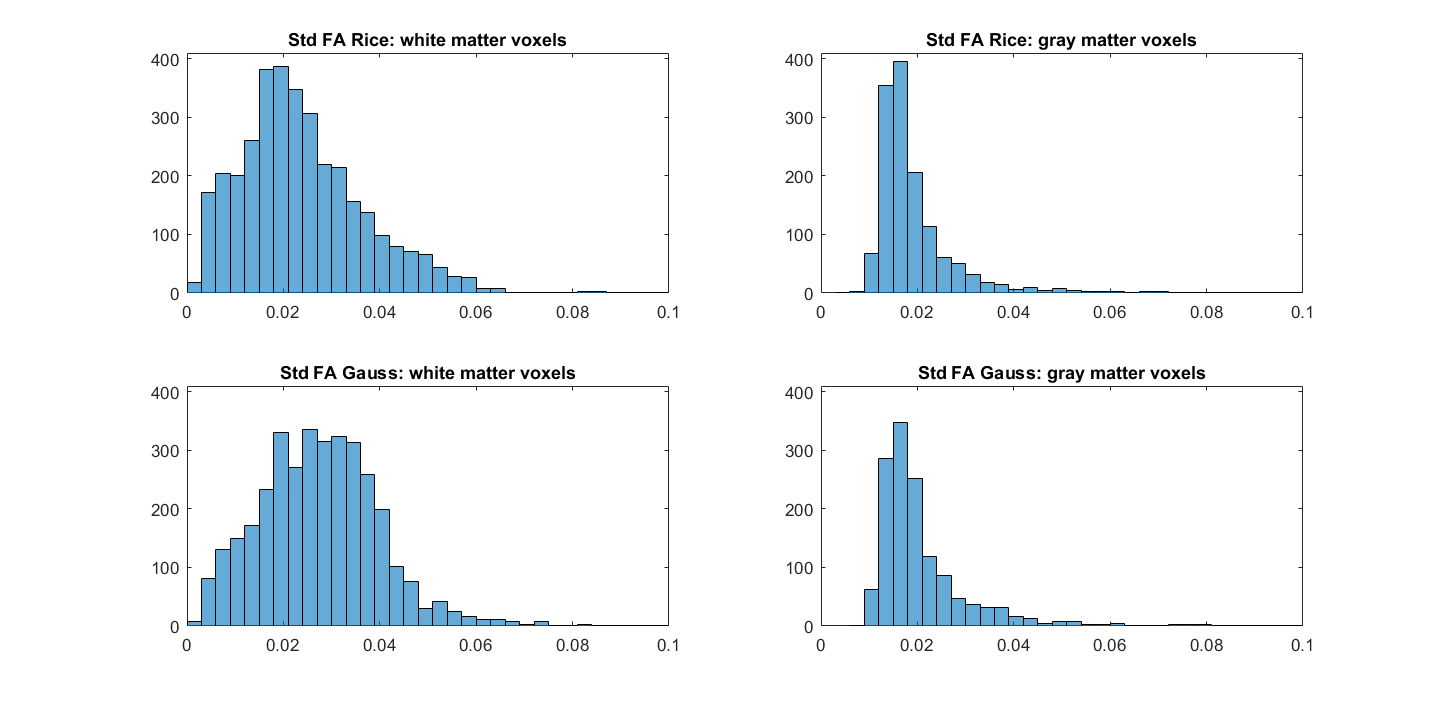}
\par\end{centering}
\caption{Histograms of posterior standard deviations of FA for white-matter
(left) and gray-matter (right) voxels for the Rician and Gaussian
DTI models, using the whole dataset. A white-matter (gray-matter)
voxel is defined as a voxel where the probability is 1 for white matter
(gray matter) from the function FAST in FSL.\label{fig:WhiteGrayStdFA}}
\end{figure}

\clearpage

\section{Discussion}

We propose a Bayesian non-central $\chi$ regression model for neuroimaging
with the Rician model as a prominent special case. The model is applied
to real diffusion data from the Human Connectome Project \citep{VanEssen}
and to simulated fMRI data with different SNRs. We show that the results
from the theoretically correct Rician DTI model can differ substantially
from the approximate Gaussian model typically used for diffusion tensor
estimation. The Gaussian model greatly underestimates the mean diffusivity
(MD) and substantially underestimates the FA of the single-diffusion
tensors, which is consistent with previous results \citep{andersson}.
We also show that the differences between the Rician and Gaussian
models increase with the b-value, which is natural since the SNR decreases
with a higher b-value. Our results for real fMRI datasets are consistent
with previous work \citep{solo2007EMRician,AdrianMaitraRowe2013Ricean},
which also come to the conclusion that there are negligible differences
between the Rician and Gaussian noise models. We demonstrate, however,
that the Rician model is remarkably adept at recovering the activations
for simulated fMRI datasets at very low SNRs, which are more common
in high-resolution images; we also show that the Gaussian model fails
to detect activity for low SNRs.

Our framework is more general compared to the work by \citet{andersson}
and other frameworks, as it is possible to include covariates for
both the mean and the variance of the noise, and not only covariates
for the mean. We show that DTI noise of the underlying complex-valued
signal is heteroscedastic, especially for the Gaussian model. This
is consistent to our recent work \citep{WegmannHetero}, where we
showed that using diffusion covariates for the noise variance gives
rather different results for DTI. It is also possible to include head
motion parameters, and their temporal derivatives, as covariates for
the noise variance for both fMRI and DTI. This can for example be
used to down-weight measurements close to motion spikes \citep{Power2014,Elhabian,siegel}
(as any measurement with a high variance is automatically down-weighted
in our framework). For models with a large number of covariates, our
variable selection can automatically discard covariates of no interest.

A potential drawback of our approach is the computational complexity.
It takes 5.6 seconds to run 1,000 MCMC iterations for the Gaussian
model in a representative voxel for the DTI data, and 11.2 seconds
for the Rician model. For a typical DTI dataset with 20,000 brain
voxels, this gives a total processing time of 31.1 hours for the Gaussian
model and 62.2 hours for the Rician model. For this reason, we have
only analyzed a single subject, as a group analysis with 20 subjects
would be rather time consuming. As each voxel is analyzed independently,
it is in theory straightforward to run MCMC on the voxels in parallel,
using a CPU or a GPU \citep{Eklund,Guo}.

We have focused on the rather simple single-diffusion tensor, while
more recent work focus on extending the diffusion tensor to higher
orders. In the work by \citet{Westin}, a regression approach is used
to estimate the diffusion tensor and a fourth order covariance matrix
in every voxel. Our regression framework can therefore easily be applied
to QTI (q-space trajectory imaging) data \citep{Westin} as well,
and more generally for any diffusion model that can be estimated using
regression. As a fourth order covariance matrix contains 21 independent
variables, the possibility to perform variable selection becomes even
more important. Furthermore, DTI is still the most common choice for
studies investigating FA differences between healthy controls and
subjects with some disease \citep{Shenton,Eierud}. Another indicator
of the importance of FA is that the TBSS approach \citep{Smith} has
received more than 2,800 citations (with about 500 citations in 2015).
Our approach gives the full posterior distribution of the FA, and
any other function of the diffusion tensor, which can be used for
tractography and to down-weight subjects with a higher uncertainty
in a group analysis. This is in contrast to TBSS and the work by \citet{andersson},
which ignore the uncertainty of the FA. \citet{andersson} develops
a sophisticated maximum a posteriori (MAP) estimation for the DTI
model, but does not deal with posterior uncertainty, in contrast to
our full MCMC sampling from the posterior distribution.

\section*{Acknowledgement}

Anders Eklund was supported by the Information Technology for European
Advancement (ITEA) 3 Project BENEFIT (better effectiveness and efficiency
by measuring and modelling of interventional therapy) and by the Swedish
research council (grant 2015-05356, ``Learning of sets of diffusion
MRI sequences for optimal imaging of micro structures``). Anders
Eklund and Bertil Wegmann were supported by the Swedish research council
(grant 2013-5229, ``Statistical analysis of fMRI data'').

Data collection and sharing for this project was provided by the Human
Connectome Project (HCP; Principal Investigators: Bruce Rosen, M.D.,
Ph.D., Arthur W. Toga, Ph.D., Van J. Weeden, MD). HCP funding was
provided by the National Institute of Dental and Craniofacial Research
(NIDCR), the National Institute of Mental Health (NIMH), and the National
Institute of Neurological Disorders and Stroke (NINDS). HCP data are
disseminated by the Laboratory of Neuro Imaging at the University
of Southern California.

\appendix

\section{Gradients and Hessians \label{Appendix A}}

\citet{Villani2012GSM} derive the gradient and the Hessian for a
general posterior of the form
\begin{equation}
p(\beta|\mathbf{y},\mathbf{x})\propto\prod_{i=1}^{n}p(y_{i}|\phi_{i},\mathbf{x}_{i})p(\beta),\label{eq:GeneralPosterior}
\end{equation}
where $k(\phi_{i})=\mathbf{x}_{i}^{\prime}\beta$ is a smooth link
function, and $\mathbf{x}_{i}$ is a covariate vector for the $i$th
observation; the full conditional posteriors for $\beta$ and $\alpha$
in the Rician and the NC-$\chi$ case with logarithmic links are clearly
of this form. The gradient of the likelihood in Eq. (\ref{eq:GeneralPosterior})
can be expressed as 
\begin{equation}
\frac{\partial\ln p(\mathbf{y}|\beta,\mathbf{X})}{\partial\beta}=\mathbf{X}^{T}\mathbf{g},\label{General Gradient}
\end{equation}
where $\mathbf{X}=(\mathbf{x}_{1},...,\mathbf{x}_{n})^{T}$, $\mathbf{g}=(g_{1},...,g_{n})^{\prime}$,
and
\[
g_{i}=\frac{\partial\ln p(y_{i}|\phi_{i})}{\partial\phi_{i}}\left[k^{\prime}(\phi_{i})\right]^{-1}.
\]
The Hessian of the likelihood is
\begin{equation}
\frac{\partial^{2}\ln p(\mathbf{y}|\mathbf{X},\beta)}{\partial\beta\partial\beta^{\prime}}=\mathbf{X}^{T}\left(\mathbf{D}_{1}+\mathbf{D}_{2}\right)\mathbf{X},\label{General Hessian}
\end{equation}
where $\mathbf{D}_{1}=\mathrm{Diag}(d_{1i})$, $\mathbf{D}_{2}=\mathrm{Diag}(d_{2i})$,
\[
d_{1i}=\frac{\partial^{2}\ln p(y_{i}|\phi_{i},\mathbf{x}_{i})}{\partial\phi_{i}^{2}}\left[k^{\prime}(\phi_{i})\right]^{-2},
\]
and
\[
d_{2i}=-\frac{\partial\ln p(y_{i}|\phi_{i},\mathbf{x}_{i})}{\partial\phi_{i}}k^{\prime\prime}[k^{\prime}(\phi_{i})^{-1}]k^{\prime}(\phi_{i})^{-2}.
\]
The outer-product approximation of the Hessian is given by
\[
\mathbf{X}^{T}\mathrm{Diag}(g_{i}^{2})\mathbf{X},
\]
which is faster to compute and often numerically more stable than
the Hessian itself. Finally, the Fisher information is given by
\[
E_{\mathbf{y}|\mathbf{X},\beta}\left(\frac{\partial^{2}\ln p(\mathbf{y}|\mathbf{X},\beta)}{\partial\beta\partial\beta^{\prime}}\right)=\mathbf{X}^{T}\left[\mathrm{Diag}(E_{y_{i|\phi_{i}}}d_{1i})\right]\mathbf{X}.
\]

From \eqref{General Gradient} and \eqref{General Hessian} it is
sufficient for our MCMC algorithm to compute the first and second
derivatives with respect to (w.r.t.) $\mu$ and $\phi$, respectively,
for each of the individual observation. The log-likelihood for one
observation $y$ of a non-central $\chi$ variable can be written
as \citep{tristan2012least}
\begin{align*}
\ln p(y|\mu,\phi,L) & =L\ln y-\ln\phi-(L-1)\ln\mu-\frac{\left(y^{2}+\mu^{2}\right)}{2\phi}\\
 & +\ln I_{L-1}\left(\frac{y\mu}{\phi}\right),
\end{align*}
where $I_{L-1}(\cdot)$ is the modified Bessel function of the first
kind with order $L-1$. The following derivatives hold for $I_{L}(z)$
\begin{align*}
\frac{\partial}{\partial z}I_{0}(z) & =I_{1}(z)\\
\frac{\partial}{\partial z}I_{L}(z) & =\frac{I_{L-1}(z)+I_{L+1}(z)}{2},\;\text{for}\;L\geq1.
\end{align*}
Let $z=\frac{y\mu}{\phi}$ and 
\[
B(z)=\frac{I_{1}(z)}{I_{0}(z)},\;\text{if}\;L=1
\]
\[
B(z)=\frac{I_{L-2}(z)+I_{L}(z)}{2I_{L-1}(z)},\;\text{if}\;L\geq2
\]
Then, 
\[
B^{\prime}(z)=\frac{dB(z)}{dz}=\frac{1}{2}\left(1+\frac{I_{2}(z)}{I_{0}(z)}\right)-B^{2}(z),\;\text{if}\;L=1
\]
\[
B^{\prime}(z)=\frac{dB(z)}{dz}=\frac{1}{4}\left(3+\frac{I_{3}(z)}{I_{1}(z)}\right)-B^{2}(z),\;\text{if}\;L=2
\]
\[
B^{\prime}(z)=\frac{dB(z)}{dz}=\frac{1}{4}\left(2+\frac{I_{L-3}(z)+I_{L+1}(z)}{I_{L-1}(z)}\right)-B^{2}(z),\;\text{if}\;L\geq3.
\]

\subsection{Derivatives w.r.t. to $\mu$:}

\[
\frac{\partial\ln p(y|\mu,\phi,L)}{\partial\mu}=\frac{yB(z)-\mu}{\phi}-\frac{L-1}{\mu}
\]

\[
\frac{\partial^{2}\ln p(y|\mu,\phi,L)}{\partial\mu^{2}}=\left(\frac{y}{\phi}\right)^{2}B^{\prime}(z)-\frac{1}{\phi}+\frac{L-1}{\mu^{2}}
\]

\subsection{Derivatives w.r.t. to $\phi$:}

\[
\frac{\partial\ln p(y|\mu,\phi,L)}{\partial\phi}=\frac{1}{2}\left(\frac{y^{2}+\mu^{2}}{\phi^{2}}\right)-\frac{1}{\phi}\left(1+zB(z)\right)
\]

\begin{align*}
\frac{\partial^{2}\ln p(y|\mu,\phi,L)}{\partial\phi^{2}} & =\frac{z\left[B(z)+zB^{\prime}(z)\right]-\frac{1}{2}\left(\frac{y^{2}+\mu^{2}}{\phi}\right)}{\phi^{2}}\\
 & -\frac{\partial\ln p(y|\mu,\phi,L)}{\partial\phi}\phi^{-1}
\end{align*}
\begin{comment}

\subsection{Som jag skrev tidigare i texten i kapitel 2.2, så tycker jag att
vi kan skippa dessa derivator för $L$ som endast är för heltal $L$.
Observera även att $K\left(\cdot\right)$ är den modifierade Bessel
funktionen av den andra ordningen. Jag har låtit det stå kvar för
NC-chi modellen i Appendix A.1 och A.2 så länge.}

\subsection{Derivatives w.r.t. to $L$:}

\[
\frac{\partial\ln p(y|\mu,\phi,L)}{\partial L}=\ln\frac{y}{\mu}+\frac{I_{L-1}^{\prime}(z)}{I_{L-1}\left(z\right)}
\]

\[
\frac{\partial\ln p(y|\mu,\phi,L)}{\partial L^{2}}=\frac{I_{L-1}^{\prime\prime}(z)}{I_{L-1}(z)}-\left(\frac{I_{L-1}^{\prime}(z)}{I_{L-1}\left(z\right)}\right)^{2},
\]
where 
\[
I_{L-1}^{\prime}(z)=\frac{\partial I_{L-1}(z)}{\partial L}=(\text{\textminus}1)^{L}K_{L-1}(z),\;\text{if}\;L\leq2
\]
\[
I_{L-1}^{\prime}(z)=\frac{\partial I_{L-1}(z)}{\partial L}=(\text{\textminus}1)^{L}K_{L-1}(z)+\left(L-1\right)!\sum_{k=0}^{L-2}\frac{\left(-z\right)^{-\left(L-k-1\right)}}{k!\left(L-k-1\right)\left(2^{-\left(L-k-2\right)}\right)}I_{k}(z)\;\text{if}\;L\geq3
\]

\[
I_{L-1}^{\prime\prime}(z)=\frac{\partial I_{L-1}(z)}{\partial L^{2}}=
\]
\end{comment}

\bibliographystyle{apalike}
\bibliography{DTI}

\end{document}